\documentclass[preprintnumbers,reprint,superscriptaddress,twocolumn]{revtex4}
\usepackage{textcomp}
\usepackage{graphicx}
\usepackage{amssymb}
\usepackage{color}

\makeatletter

\begin{document}
\global\long\def\tbtb{\bar{t}\bar{t}}
\global\long\def\met{\not{\!{\rm E}}_{T}}

\title{Color Sextet Scalars in Early LHC Experiments}
\author{Edmond L. Berger}
\affiliation{High Energy Division, Argonne National Laboratory, Argonne, Illinois
60439, USA}

\author{Qing-Hong Cao}
\affiliation{High Energy Division, Argonne National Laboratory, Argonne, Illinois
60439, USA}
\affiliation{Enrico Fermi Institute, University of Chicago, Chicago, Illinois
60637, USA}

\author{Chuan-Ren Chen}
\affiliation{Institute for Physics and Mathematics of the Universe, University
of Tokyo, Chiba 277-8568, Japan}

\author{Gabe Shaughnessy}
\affiliation{High Energy Division, Argonne National Laboratory, Argonne, Illinois
60439, USA}
\affiliation{Department of Physics and Astronomy, Northwestern University, Evanston,
Illinois 60208, USA}

\author{Hao Zhang}
\affiliation{Enrico Fermi Institute, University of Chicago, Chicago, Illinois
60637, USA}
\affiliation{Department of Physics and State Key Laboratory of Nuclear Physics
and Technology, Peking University, Beijing 100871, China}

\begin{abstract}
We explore the potential for discovery of an exotic color
sextet scalar in same-sign top quark pair production 
in early running at the LHC.  We present the first phenomenological analysis 
at colliders of color sextet scalars with full top quark spin correlations 
included.  
We demonstrate that one can measure the scalar mass, the top quark polarization,
and confirm the scalar resonance with 1 fb$^{-1}$ of integrated luminosity.  
The top quark polarization can distinguish gauge triplet and singlet scalars.

\end{abstract}

\preprint{ANL-HEP-PR-10-13, EFI-10-11, IPMU10-0076, NWU-HEP-TH/10-07} 

\maketitle

\noindent{\bf Introduction --} The Large Hadron Collider (LHC) at CERN 
opens a new energy frontier in searches for new physics beyond the 
standard model (SM) of particle physics.  
According to the current LHC operation schedule,  a data sample based on 
$\sim 1~\rm{fb}^{-1}$ of integrated luminosity at 7 TeV center of mass energy 
is anticipated within two years.  Signs of new physics (NP) would be observed 
at this luminosity if  (i) the collider signature is novel and easily detectable;
(ii) the cross section is relatively large;  
and (iii) the SM backgrounds are suppressed.    

In this Letter we present an analysis of a color sextet
scalar model which satisfies all three key points, 
and we discuss the requirements for discovery and verification of the model.   
The final state of special interest to us is a {\em same-sign} top quark pair.  
We show that the mass of the new colored scalar can be estimated well and 
that its scalar nature can be verified by a measurement of the spin correlations 
of the final leptons from the decay of the top quarks.  Among other 
consequences, observation of color sextet scalars would alter the evolution 
of the strong coupling strength and indicate a fundamentally 
different picture for unification from conventional grand unified theories. 
We exploit the valence quark content of the initial protons at the LHC by 
focusing on a  production mechanism in which the quark-quark interactions play an important role in the 
production of a heavy resonance for which large values of the partonic 
Bjorken-$x$ are needed, a region in which the gluon parton density drops 
off rapidly but the quark density in a proton remains large.  

A bosonic colored state can be produced in quark-quark fusion with 
color structure obtained from $3\times3=6\oplus\bar{3}$, where $3$, $6$, 
and $\bar{3}$ are the triplet, sextet,  and antitriplet representations of the 
quantum chromodynamics $SU(3)_{C}$ color group.  Color sextet 
scalars are present in partial unification~\cite{Pati:1974yy}.  
The masses of the scalars can be possibly as low as the weak 
scale, $\sim\rm{TeV}$ or less~\cite{Chacko:1998td}.
The fermion number violating interaction of a new scalar $\Phi$ with two 
quarks is~\cite{Han:2009ya} 
\begin{equation}
  \mathcal{L}_{6/\bar{3}}  =  
  \Phi_{j}^{*}K_{ab}^{j}q_{a}^{T}C^{\dagger}(\lambda_{qq^\prime}^{L} P_{L}
  +\lambda_{qq^\prime}^{R}P_{R})q^\prime_{b}+H.c.\,\,,
\end{equation}
where $K_{ab}^{j}$ is the Clebsch Gordan coefficient~\cite{Han:2009ya} of 
the 6 or $\bar{3}$ representation, 
$a$ and $b$ are the color indices of quarks, $P_{L/R}$ is the usual
left- or right-handed projector and, without loss of generality, the
Yukawa coupling $\lambda^{L/R}$ is real in this study.
The coupling $\lambda^{L/R}$ can be relatively large since it is not 
proportional to the quark mass.  
The coupling of the new scalar to two up-type quarks is constrained by the 
measurement of $D^{0}-\bar{D}^{0}$
mixing and by the rate of $D\to\pi^{+}\pi^{0}(\pi^{+}\phi)$ decay, resulting in 
stringent bounds on the coupling $\lambda_{qq^\prime}$:
$\left|{\rm Re}(\lambda_{cc}\lambda_{uu}^{*})\right|\sim5.76\times10^{-7}$
for $m_\Phi \sim 1~{\rm TeV}$~\cite{Chen:2009xjb}.  However, this strong constraint 
can be relaxed if the coupling to the second generation quarks is minimized.  
We concentrate on the flavor conserving couplings 
and treat $\lambda_{uu}$ and $\lambda_{tt}$ as independent couplings.   Alternatively, the 
first generation may be suppressed while the second generation is not, but such sea-quark 
initiated processes are relatively suppressed.

A color scalar can carry the following 
electroweak quantum numbers: (6,1,4/3), (6,1,-2/3), (6,1,1/3), and (6,3,1/3), 
where the numbers in the parentheses denote $(SU(3)_C, SU(2)_L,U(1)_Y)$;
see Ref.~\cite{Cakir:2005iw,Han:2009ya,Arnold:2009ay} for details.   
The SM gauge symmetry demands that $\Phi_{6,1,4/3}$ interacts with $u_R u_R$, $\Phi_{6,1,-2/3}$ with $d_R d_R$, $\Phi_{6,1,1/3}$ with $u_L d_L$ and $u_R d_R$, and $\Phi_{6,3,1/3}$ with $(\sqrt{2} u_L u_L, -u_L d_L, \sqrt{2} d_L d_L)$~\cite{Cakir:2005iw}, where $(u_L,d_L)$ denotes the $SU(2)_L$ quark doublet while $u_R$ and $d_R$ are the corresponding right-handed singlet fields.  A measurement of the spin of quarks from $\Phi$ decay would decipher the handedness of the coupling of the scalar to quarks and also the gauge quantum number of $\Phi$.  With this goal in mind, we consider the decay of $\Phi \to tt$ in this work because a top quark decays promptly via the weak interaction such that its spin information is retained among its decay products.
We emphasize that top quark polarization can be used to distinguish gauge triplet
from gauge singlet scalars because triplet scalars decay to left-handed top quarks
while singlet scalars decay to right-handed top quarks. We concentrate here on the specific case of $\Phi_{6,1,4/3}$ which couples only to right-handed gauge singlet up-type quarks, but our work can be easily extended to include a study of $\Phi_{6,3,1/3}$.  

The most striking collider signature of a color sextet scalar is same-sign top quark pair 
production in a direct $s$-channel process, as is pointed out in Refs.~\cite{Mohapatra:2007af} 
where top quark decay is not considered.  
It is important to retain the top quark spin correlations in the calculation
because the kinematic distributions of the decay products depend strongly on the
top quark spin.  The new scalar interacts only with right-handed gauge singlet quarks so 
the top quarks from its decay are mainly right-handed.  A charged lepton 
from a right-handed top quark decay has a harder momentum spectrum than one from a left-handed top 
quark.   No available event generator calculates color sextet scalar induced $tt$ production and subsequent $t$ decay with full top quark spin correlations included.   We calculate the helicity amplitudes for $uu \to tt \to bW^+(\to \ell^+ \nu) bW^+(\to \ell^+ \nu)$ and implement the process in our own parton-level Monte Carlo code. 

Because the decay width of $\Phi$ is narrow, $ \Gamma(\Phi\to qq) \approx \frac{1}{16\pi} m_{\Phi} \lambda_{qq}^2$, one can factor the process $uu\to tt$ into scalar production and decay 
terms,  
\begin{eqnarray}
&& \sigma(uu\to\Phi\to tt) =\sigma_{0}(uu\to\Phi) \times \lambda_{uu}^{2} {\rm Br}(tt), 
\nonumber \\
 &=& \sigma_{0}(uu\to\Phi\to tt) \times \lambda_{uu}^{2} \frac{{\rm Br}(tt)}{{\rm Br}_0(tt)}.
        \label{eq:conv}
\end{eqnarray}
The decay branching ratio ${\rm Br}(tt)\equiv{\rm Br}(\Phi\to tt)$ is 
\begin{equation}
{\rm Br}(tt)=\frac{\lambda_{tt}^2 R}{\lambda_{uu}^2+\lambda_{tt}^2 R},
R\equiv \sqrt{1-\frac{4 m_t^2}{m_\Phi^2}}\left( 1-\frac{2m_t^2}{m_\Phi^2}\right).
\end{equation}
Subscript ``0" denotes $\lambda_{uu}=\lambda_{tt}=1$.  
We choose to work with the following two parameters
in the rest of this paper: the scalar mass $m_\Phi$ and the product 
$\lambda_{uu}^2 {\rm Br}(\Phi\to tt)$.  The kinematics of the final state 
particles are determined by the scalar mass, whereas the couplings of the 
scalar to the light and heavy fermions change the overall normalization. 

\begin{figure}
\includegraphics[clip,scale=0.43]{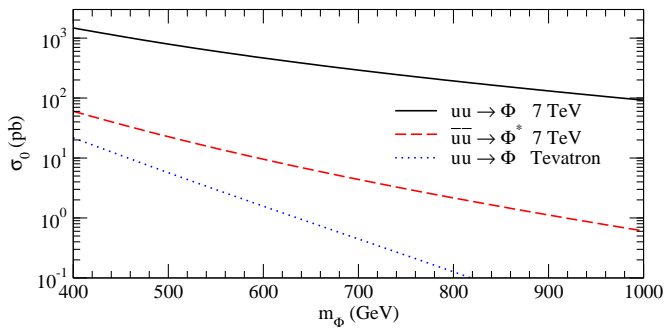}
\caption{Leading order cross sections (pb) via $uu\to \Phi$ and $\bar{u}\bar{u}\to\Phi^{*}$ 
at the Tevatron (dotted) and at the 7~TeV LHC (solid and dashed) for $\lambda_{uu}=1$.  We use the CTEQ6L parton distribution functions~\cite{Pumplin:2002vw} and choose the renormalization and factorization scales as $m_\Phi$.
\label{fig:xsect} }
\end{figure}

Figure~\ref{fig:xsect} displays the $\Phi$ production cross sections via
$uu\to\Phi$ and $\bar{u}\bar{u}\to \Phi^{*}$ at the Tevatron (dotted)
collider and at the 7~TeV LHC.  
The $uu$ initial state dominates $\bar{u}\bar{u}$ for a $pp$ initial state.  
Therefore, $tt$ pairs are produced much more than $\tbtb$ pairs,  
leading to an asymmetry of the same-sign charged leptons. 
This doubly positive excess would be an early hint of a color sextet scalar.   

\begin{figure}[t]
\includegraphics[clip,scale=0.48]{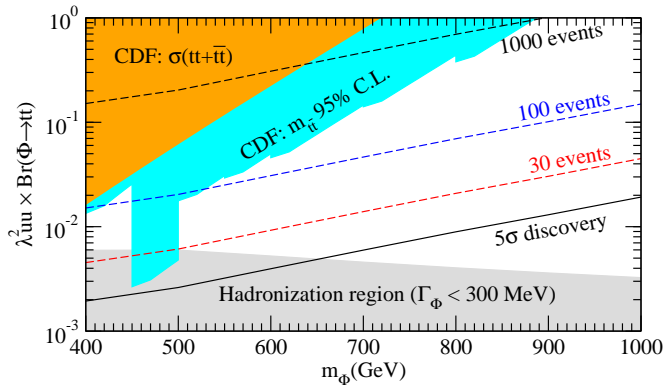}
\caption{Discovery potential of the process $uu\to\Phi\to tt$ at the 7~TeV
LHC with ${\cal L}=1\text{ fb}^{-1}$. The shaded regions are excluded by the Tevatron measurements:
the orange band is derived from the search of same-sign top quark pair production; 
the cyan band denotes the 95\% C.L. constraints
from the $m_{t\bar t}$ measurements.  The gray region indicates that $\Phi$ would hadronize before decaying, washing out spin-correlations.
\label{fig:disovery-7tev} }
\end{figure}

The search for the same-sign top quark pair production in the dilepton 
mode at the Tevatron imposes an upper limit 
$\sigma(tt+\tbtb)\leq0.7\,{\rm pb}$~\cite{Amsler:2008zzb}. 
The CDF collaboration measured the $t\bar{t}$ invariant mass
spectrum in the semileptonic decay mode~\cite{Aaltonen:2009iz}.  Since $b$ and 
$\bar{b}$ jets from $t \rightarrow W b$ are not distinguished well, $tt$ pairs 
lead to the same signature as $t\bar{t}$ in the semileptonic mode.  Hence, 
the $m_{t\bar{t}}$ spectrum provides an upper limit on $\sigma(tt+\tbtb)$,
shown in 
the cyan shaded region in Fig.~\ref{fig:disovery-7tev}.  The lower gray 
shaded region is the region in which $\Phi$ would 
hadronize before decay, washing out the spin correlation effects we 
utilize to probe the coupling and spin of the sextet state.

\noindent{\bf Event simulation --}  We focus on 
the same-sign dilepton decay mode, in which the $W$ bosons from both $t \rightarrow W b$ decays 
lead to a final state containing an electron or muon, $W \rightarrow l \nu$, accounting for 
about $5\%$ of all $tt$ decays.
We concentrate on the clean $\mu^{+}\mu^{+}$ final state because
the muon reconstruction has a large average efficiency of $95-99\%$ within the  
pseudorapidity range $\left|\eta\right|<2.4$ and transverse momentum range 
$5\text{ GeV}\le p_{T}\le 1\text{ TeV}$, 
while the charge misassigned fraction for muons with $p_{T}=100\,{\rm GeV}$
is less than $0.1\%$~\cite{Aad:2009wy}. 

These events are characterized by two high-energy same-sign leptons,
two jets from the hadronization of the $b$-quarks, and large missing
energy ($\not{\!\!{\rm E}}_{T}$) from two unobserved neutrinos. The dominant backgrounds
yielding the same collider signature are the processes 
(generated here with ALPGEN~\cite{Mangano:2002ea}):
\begin{eqnarray}
pp&\to& W^{+}(\to \ell^+\nu)W^{+}(\to \ell^+\nu)jj, \\
pp&\to& t\bar{t} \to bW^{+}(\to \ell^+\nu)\bar{b}(\to \ell^+)W^{-}(\to jj).
\end{eqnarray}
The first process ($WWjj$) is the SM irreducible background
while the second ($t\bar{t}$) is a reducible background as it contributes
when some tagged particles escape detection, carrying small
$p_{T}$ or falling out of the detector rapidity coverage.  For example,
one of the $b$-quarks
decays into an isolated charged lepton while one of the two jets from the $W^{-}$
boson decay is mistagged as a $b$-jet. 
Other SM backgrounds, e.g. triple gauge boson production ($WWW$, $ZWW$, and $WZg(\to b\bar{b})$), occur at a negligible rate after kinematic cuts, and are not shown here.  

At the analysis level, all signal and background events are required to pass the
acceptance cuts $p_{T,\,j},p_{T,\,\ell}\geq50\,{\rm GeV}$, $\left|\eta_{j}\right|\leq2.5$, $\left|\eta_{\ell}\right|\leq2.0$, and $\Delta R_{jj,j\ell,\ell\ell} > 0.4$,  
where the separation in the azimuthal angle ($\phi$)-pseudorapidity ($\eta$) plane between
the objects $k$ and $l$ is 
$\Delta R_{kl}\equiv\sqrt{\left(\eta_{k}-\eta_{l}\right)^{2}+\left(\phi_{k}-\phi_{l}\right)^{2}}$.
We model detector resolution effects by smearing the final
state energy according to 
$\delta E/E= \mathcal{A}/\sqrt{E/{\rm GeV}}\oplus \mathcal{B}$,
where we take $\mathcal{A}=10(50)\%$ and $\mathcal{B}=0.7(3)\%$ for leptons(jets). To
account for $b$-jet tagging efficiencies, we demand two $b$-tagged jets, each with
a tagging efficiency of $60\%$.
We also apply a mistagging rate for charm-quarks 
$\epsilon_{c\to b}=10\%$ for $p_{T}(c)>50\,{\rm GeV}$.
The mistag rate for a light jet is
$\epsilon_{u,d,s,g\to b}= 0.67\%$ for $p_{T}(j)<100\,{\rm GeV}$
and $2\% $ for $p_{T}(j)>250\,{\rm GeV}$.
For $100\,{\rm GeV}<p_{T}\left(j\right)<250\,{\rm GeV}$,
we linearly interpolate the fake rates given above~\cite{Berger:2009cm}. 

\begin{table*}
\caption{Signal and background cross sections (pb) before and after cuts, with $\lambda_{uu} =\lambda_{tt} =1$, for six values of $m_{\Phi}$~(GeV).   The decay branching ratios of the 
signal ${\rm Br}(tt)$ are given in the second column.   The ``no cut'' rates correspond to all lepton 
and quark decay modes of $W$-bosons, whereas those  ``with cut''  are obtained after all cuts, the 
restriction to 2 $\mu^+$'s and with tagging efficiencies included.   
\label{tab:xsec}}
\begin{tabular}{cccc|cccc|ccccc}
\hline 
$m_{\Phi}$ & ${\rm Br}(tt)$  & No cut  &  With cut & $m_{\Phi}$ & ${\rm Br}(tt)$  & No cut  &  With cut &Background & No cut & With cut\tabularnewline
\hline
500  & 0.35  & 288.44   & 1.71 & 800  & 0.45  &  91.04  & 0.65  & $t\bar{t}$  & 97.62   & 0.0032\tabularnewline
600  & 0.41  & 193.67   & 1.30 & 900  & 0.46  &  65.14  & 0.45  & $WWjj$      &  9.38   & 0.0014\tabularnewline
700  & 0.43  & 133.46   & 0.93 & 1000 & 0.47  &  46.72  & 0.31  & $WWW/Z$     &  0.03   & 0\tabularnewline
\hline
\end{tabular}
\end{table*}

After lepton and jet reconstruction, we demand two hard leptons of the same sign, a requirement 
which greatly reduces the SM background, giving a rejection of order $10^{-4}$ and $10^{-3}$ for 
the $t\bar t$ and $WWjj$ processes, respectively.  After the cuts are imposed, we find a total of 4.6 background events, 3.2 from $t\bar t$ and 1.4 from $WWjj$ for 1 fb$^{-1}$ of integrated luminosity.  After $b$-tagging and restriction to the $\mu^+\mu^+$ mode, more than half the signal events survive the analysis cuts.  Signal and background cross sections are shown 
in Table~\ref{tab:xsec}, before and after cuts, for 6 values of $m_{\Phi}$.

In Fig.~\ref{fig:disovery-7tev}, we show the expected numbers of signal events as a function of $m_\Phi$ for a range of values of the coupling $\lambda_{uu}^2 \text{Br}(\Phi\to tt)$.  We obtain the event rate lines by converting the required cross section into $\lambda_{uu}^2 \text{Br}(\Phi\to tt)$ via Eq.~\ref{eq:conv}.  In a similar fashion, the $5\sigma$ discovery line is found by requiring 14 signal events.
Rates for other values of $\lambda_{uu}$ and $\lambda_{tt}$ can be obtained from Eq.~\ref{eq:conv}.

\noindent{\bf Reconstruction and Measurements --}   Given a sample of events having two like-sign muons, two $b$-tagged jets, and missing energy, we now address the possibility of establishing that these events are consistent with $\Phi \to tt \to bW^+(\to \ell^+ \nu) bW^+(\to \ell^+ \nu)$.   
A full event reconstruction is required.  We demonstrate that the $M_{T2}$ variable can be used for full kinematic reconstruction and to identify the $W$-bosons and top quarks in the intermediate states.
 
The $M_{T2}$ event variable is designed to bound the masses of a
pair of heavy particles that subsequently decay into one or more visible states and missing energy. It is 
a function of the momenta of two visible particles and the missing transverse momentum in an event~\cite{Lester:1999tx}.  Strictly speaking, $M_{T2}(a,b,\met)$ is the minimum of a function
\begin{equation}
\max \left\{ M_T(\vec{p}_T^{~a},\not{\! p_1}),M_T(\vec{p}_T^{~b},\not{\! p_2})\right\} ,
\end{equation}
such that $\not{\!p}_{1,T}+  \not{\!p}_{2,T}=\met$.  Here $a$ and $b$ are the two individual (or clustered) visible states from the parent decay, and $\not{\!p}_1, \not{\!p}_2$ are the associated missing momenta.  
The transverse mass $M_{T}$ is defined as 
\begin{equation}
M_{T}(X,p_T^{\rm invis}) = \sqrt{m_X^2 + 2 ( E_T^X E_T^{\rm invis} - \vec{p}_T^X \cdot \vec{p}_{T}^{\rm~invis}) },
\end{equation}
where $X$ denotes the visible particle or cluster. 

The $M_{T2}$ approach requires the correct assignment of final state particles to the $a$ and $b$ clusters if the parent particle is to be reconstructed successfully.  
For the $W$-boson, the assignment is straightforward (one $\mu^+\to a$ and the other $\mu^+\to b$), and it yields an endpoint at the $W$-boson mass, confirming the two $\mu^+$ leptons and $\met$ originate from the decay of $W$-bosons.  However, there are two possible combinations of $\mu$-$b$ clusters for reconstructing the $t$-quark mass.  To choose the correct combination, we require the $\mu$-$b$ pairing to be such that $M_{T2}$ is minimized.   This choice picks the correct combination with $95\%$ probability.  The same method is used successfully in Ref.~\cite{Aaltonen:2009rm} to determine the top quark mass from a $t \bar{t}$ sample, with both $t$ quarks decaying leptonically.  

\begin{figure}[t]
\includegraphics[clip,scale=0.6]{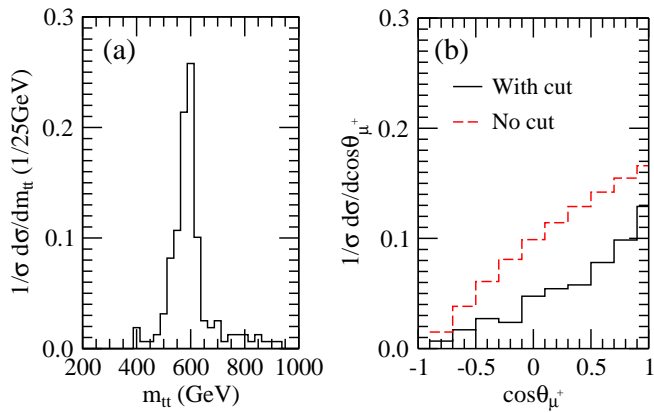}
\caption{(a) Normalized $tt$ invariant mass distribution for 
$m_\Phi=600$ GeV and $\lambda_{uu}=\lambda_{tt}=0.3$ after the neutrino momenta are reconstructed.  
(b) Distribution of the angle of the charged lepton relative to the top quark in the c.m. frame before (red dotted) and after (black) cuts and efficiencies are included.  Separation cuts distort the distribution only slightly, allowing one to read the top quark polarization with few events.  Distributions shown for 1 fb$^{-1}$ of simulated data for the signal
only;  the SM background is negligible. }
\label{fig:toppol}
\end{figure}

Once the correct pairing of $b$ and $\mu^+$ is made, the entire event can be reconstructed.
There are 6 unknowns from $\not{\!p}_{1}$ and $\not{\!p}_{2}$, two of which 
are eliminated by the $\met$ measurement. The 4 unknowns may be resolved
via the four on-shell conditions of $W$ bosons and top quarks, 
resulting in a four-fold solution~\cite{Sonnenschein:2006ud}.  Two of these solutions are real, providing momentum determination of the two neutrinos.  The top quarks from the $\Phi$ decay may therefore be fully reconstructed, allowing the mass of $\Phi$ to be determined via the $tt$ invariant mass, as shown in Fig.~\ref{fig:toppol}a.

Given the top quark four-momenta, the spin of $\Phi$ could in principle be found via the polar angle distribution of the top quark in the reconstructed rest frame of $\Phi$.  However, distinguishing the spin-0 case from a spin-1 can be difficult with limited statistics.  Instead, one may probe the coupling structure and spin of the $\Phi$ via the polarization of the top quarks.  Among the decay products of a top quark, the charged lepton is well correlated with the top quark spin direction~\cite{Mahlon:1995zn}.  In the helicity basis, the polarization of the top quark
can be found from the distribution in $\theta_{\rm hel}$, the angle of the lepton in the rest frame of top quark relative to the top quark direction of motion in the overall c.m. frame.  The angular correlation of the lepton $\ell^+$ is given by ${1\over 2}(1\pm\cos\theta_{\rm hel})$,  with the ($+$) choice for right-handed and ($-$) for left-handed top-quarks.  In Fig.~\ref{fig:toppol}b, we present this normalized distribution for a right-handed top quark before cuts (red) and after cuts (black) with 1 fb$^{-1}$ of simulated data for a scalar mass of $m_\Phi=600$ GeV.  The cuts produce some distortion, but the linear dependence on $\cos\theta_{\rm hel}$ is robust and certainly allows discrimination from ${1\over 2}(1 - \cos\theta_{\rm hel})$.  For instance, the $\cos\theta_{\mu^+}$ distribution after cuts peaks in the direction of the reconstructed top quark, $\cos \theta_{\rm hel} \sim 1$, since the hard lepton $p_T$ from the top quark decay more easily passes the lepton threshold cuts.  Based on Poisson statistics, the distribution in  Fig.~\ref{fig:toppol}b can be distinguished from an unpolarized flat distribution with as few as ${\cal O}(30)$ events at 95\% C.L. spread over 3 bins, making this measurement possible in an early LHC scenario.  

If both top quarks exhibit the same polarization in the rest frame of $\Phi$, either $t_L t_L$ or $t_R t_R$, one can conclude the resonant $\Phi$ state is a scalar rather than a vector, as a color sextet vector would decay into $t_L t_R$~\cite{Cakir:2005iw}.  The $SU(2)_L$ quantum number of $\Phi$ is also uniquely determined: a triplet for $t_L t_L$ while a singlet for $t_R t_R$. 

\noindent{\bf Acknowledgments --} The work by E.L.B., Q.H.C. and G.S. is supported in part by the U.S. DOE under Grants No.~DE-AC02-06CH11357. Q.H.C is also supported in part
by the Argonne National Laboratory and University of Chicago Joint
Theory Institute Grant 03921-07-137. C.R.C. is supported by
World Premier International Initiative, 
MEXT, Japan.  G.S. is also supported in part by the U.S. DOE under Grant No. DE-FG02-91ER40684.  H.Z. is supported in part by the National Natural Science Foundation of China under
Grants 10975004 and the China Scholarship Council File No. 2009601282.


\begin{thebibliography}{40}

\bibitem{Pati:1974yy}
  J.~C.~Pati and A.~Salam,
  Phys.\ Rev.\  D {\bf 10}, 275 (1974)
  [Erratum-ibid.\  D {\bf 11}, 703 (1975)];
  R.~N.~Mohapatra and R.~E.~Marshak,
  Phys.\ Rev.\ Lett.\  {\bf 44}, 1316 (1980)
  [Erratum-ibid.\  {\bf 44}, 1644 (1980)].

\bibitem{Chacko:1998td}
  Z.~Chacko and R.~N.~Mohapatra,
  Phys.\ Rev.\  D {\bf 59}, 055004 (1999).

\bibitem{Han:2009ya}
  T.~Han, I.~Lewis and T.~McElmurry,
  JHEP {\bf 1001}, 123 (2010).


\bibitem{Chen:2009xjb}
  C.~H.~Chen,
  Phys.\ Lett.\  B {\bf 680}, 133 (2009);
  C.~R.~Chen, W.~Klemm, V.~Rentala and K.~Wang,
  Phys.\ Rev.\  D {\bf 79}, 054002 (2009).



\bibitem{Cakir:2005iw}
  O.~Cakir and M.~Sahin,
  Phys.\ Rev.\  D {\bf 72}, 115011 (2005).
  
  
\bibitem{Arnold:2009ay}
  J.~M.~Arnold, M.~Pospelov, M.~Trott and M.~B.~Wise,
  JHEP {\bf 1001}, 073 (2010).
  
\bibitem{Mohapatra:2007af}
  R.~N.~Mohapatra, N.~Okada and H.~B.~Yu,
  Phys.\ Rev.\  D {\bf 77}, 011701 (2008).
  


\bibitem{Pumplin:2002vw}
  J.~Pumplin {\it et al.},
  JHEP {\bf 0207}, 012 (2002).

\bibitem{Amsler:2008zzb}
  C.~Amsler {\it et al.},
  Phys.\ Lett.\  B {\bf 667}, 1 (2008);
  Q.-H.~Cao {\it et al.},
  Phys.\ Rev.\  D {\bf 81}, 114004 (2010).

\bibitem{Aaltonen:2009iz}
  T.~Aaltonen {\it et al.},
  Phys.\ Rev.\ Lett.\  {\bf 102}, 222003 (2009).

\bibitem{Aad:2009wy}
  G.~Aad {\it et al.},
  arXiv:0901.0512;
  G.~L.~Bayatian {\it et al.},
  J.\ Phys.\ G {\bf 34}, 995 (2007).

\bibitem{Mangano:2002ea}
  M.~L.~Mangano {\it et al.},
  JHEP {\bf 0307}, 001 (2003).

\bibitem{Berger:2009cm}
  E.~L.~Berger, C.~B.~Jackson and G.~Shaughnessy,
  Phys.\ Rev.\  D {\bf 81}, 014014 (2010).

\bibitem{Lester:1999tx}
  C.~G.~Lester and D.~J.~Summers,
  Phys.\ Lett.\  B {\bf 463}, 99 (1999).

\bibitem{Aaltonen:2009rm}
  T.~Aaltonen {\it et al.},
  Phys.\ Rev.\  D {\bf 81}, 031102 (2010).

\bibitem{Sonnenschein:2006ud}
  L.~Sonnenschein,
  Phys.\ Rev.\  D {\bf 73}, 054015 (2006);
  M. Davids {\it et al.},  
  CMS Note {\bf 2006/077} (2006);
  Y.~Bai and Z.~Han,
  JHEP {\bf 0904}, 056 (2009).

\bibitem{Mahlon:1995zn}
  G.~Mahlon and S.~J.~Parke,
  Phys.\ Rev.\  D {\bf 53}, 4886 (1996).


\end{thebibliography}
\end{document}